\documentclass[twocolumn]{aastex631}
\usepackage{CJK}
\usepackage{amsmath}

\begin{document}
\begin{CJK*}{UTF8}{gbsn}

\title{The $M_{\rm BH}-M_\star$ Relation of the hyperluminous Dust-obscured Quasars up to $z \sim 4$}

\correspondingauthor{Yibin Luo,Lulu Fan}
\email{yibinluo@mail.ustc.edu.cn,llfan@ustc.edu.cn}

\author[0000-0002-8079-6525]{Yibin Luo (罗毅彬)}
\affiliation{Department of Astronomy, University of Science and Technology of China, Hefei 230026, China}
\affil{School of Astronomy and Space Science, University of Science and Technology of China, Hefei 230026, China}

\author[0000-0003-4200-4432]{Lulu Fan (范璐璐)}
\affiliation{Department of Astronomy, University of Science and Technology of China, Hefei 230026, China}
\affil{School of Astronomy and Space Science, University of Science and Technology of China, Hefei 230026, China}
\affil{College of Physics, Guizhou University, Guiyang 550025, China}
\affil{Deep Space Exploration Laboratory, Hefei 230088, China}

\author[0009-0004-7885-5882]{Weibin Sun (孙卫斌)}
\affiliation{Department of Astronomy, University of Science and Technology of China, Hefei 230026, China}
\affiliation{School of Astronomy and Space Science, University of Science and Technology of China, Hefei 230026, China}

\author[0009-0008-1319-498X]{Haoran Yu (于浩然)}
\affiliation{Department of Astronomy, University of Science and Technology of China, Hefei 230026, China}
\affiliation{School of Astronomy and Space Science, University of Science and Technology of China, Hefei 230026, China}

\author[0000-0002-2547-0434]{Yunkun Han (韩云坤)} 
\affiliation{Yunnan Observatories, Chinese Academy of Sciences, 396 Yangfangwang, Guandu District, Kunming, China}
\affiliation{Center for Astronomical Mega-Science, Chinese Academy of Sciences, 20A Datun Road, Chaoyang District, Beijing, 100012, China}
\affiliation{Key Laboratory for the Structure and Evolution of Celestial Objects, Chinese Academy of Sciences, 396 Yangfangwang, Guandu District, Kunming, 650216, China}
\affiliation{International Centre of Supernovae, Yunnan Key Laboratory, Kunming 650216, China}

\author[0000-0002-4742-8800]{Guangwen Chen (陈广文)}
\affiliation{Sub-department of Astrophysics, University of Oxford, Keble Road, Oxford, OX1 3RH, UK}

\author[0009-0003-5280-0755]{Mengqiu Huang (黄梦秋)}
\affiliation{Department of Astronomy, University of Science and Technology of China, Hefei 230026, China}
\affil{School of Astronomy and Space Science, University of Science and Technology of China, Hefei 230026, China}

\author[0009-0003-9423-2397]{Yihang Zhang (张迤航)}
\affiliation{Department of Astronomy, University of Science and Technology of China, Hefei 230026, China}
\affiliation{School of Astronomy and Space Science, University of Science and Technology of China, Hefei 230026, China}

\author[0000-0003-4959-1625]{Zheyu Lin (林哲宇)}
\affiliation{Department of Astronomy, University of Science and Technology of China, Hefei 230026, China}
\affil{School of Astronomy and Space Science, University of Science and Technology of China, Hefei 230026, China}

\begin{abstract}

Hot dust-obscured galaxies (Hot DOGs) are a rare population of hyperluminous dust-obscured quasars discovered by the Wide-field Infrared Survey Explorer (WISE) all-sky survey. The heavy circumnuclear dust obscuration allows only a small amount of scattered light from the obscured quasar to escape, enabling the decomposition of the stellar component from the total flux. The presence of scattered light enables the redshift of the source and the properties of the black hole to be obtained from SDSS and SDSS-related literature. From WISE and SDSS data, we select 11 hyperluminous Hot DOGs at $z=1.5-3.7$ with bolometric luminosities $L_{\rm bol} \gtrsim 10^{47}\,\mathrm{erg \ s^{-1}}$. We investigate the $M_{\rm BH}-M_\star$ relation in these sources using Bayesian spectral energy distribution (SED) fitting or with extra constraints from \textit{Hubble Space Telescope} (HST) image decomposition. Stellar masses are successfully derived for eight Hot DOGs. We find high Eddington ratios $\lambda_{\rm Edd}$ in these Hot DOGs, with the median value of 1.05 and the maximum value close to 3. The super-Eddington accretion may be associated with the overdense environments of Hot DOGs. We find no significant differences in the $M_{\rm BH}/M_\star$ of these Hot DOGs compared to the local relation, suggesting that these dust-obscured quasars are the progenitors of massive early-type galaxies. We speculate that the subsequent evolution of Hot DOGs may be significantly influenced by AGN feedback and remain on the local relation.

\end{abstract}


\section{Introduction} \label{sec:intro}
Tight correlations between the mass of supermassive black holes (SMBHs) and their host galaxy properties have been found in the local universe \citep{Magorrian1998,Haring2004,Kormendy2013}. The study of coevolution for high-redshift galaxies is currently an active research field \citep[e.g.][]{Ding2020,Ding2023,Sun2025,Tanaka2025}, with the aim of understanding when the tight correlations are established. At high redshift, studies have mostly concentrated on the relation between the SMBH masses and the total stellar masses because of various observational limitations, including the faintness and small angular sizes of high-redshift galaxies, the limited spatial resolution that prevents the structural components of the host galaxy from being resolved, and the difficulty in obtaining high-resolution, high signal-to-noise spectra for detailed kinematic analyses.

At high redshifts, BH masses are mainly estimated from single-epoch virial mass estimators based on broad AGN emission lines \citep{Vestergaard2006}. Therefore, most studies focused on type 1 AGNs \citep{Treu2007,Woo2008,Jahnke2009,Merloni2010,Schramm2013,Park2015,Suh2020,Ding2020,Ding2021b,Sun2025}.

Considering that the majority of the AGN population is obscured \citep{Hickox2018}, it is important to study the $M_{\rm BH}-M_\star$ relation of obscured AGNs. However, for obscured AGNs, it is much more difficult to measure their broad emission lines than type 1 AGNs due to dust extinction, and only a few studies has been done, focusing on red type 1 AGNs \citep{Urrutia2012} and type (1.8-1.9) AGNs \citep{Alexander2008,Del2009,Sarria2010,Melbourne2011,Bongiorno2014}.

Based on the ``W1W2-dropout" method \citep{Eisenhardt2012}, a population of hyperluminous, hot dust-obscured galaxies was discovered using the WISE and called Hot DOGs \citep{Eisenhardt2012,Wu2012}. Subsequent studies have found that Hot DOGs are hyperluminous $L_{bol} > 10^{13}L_\odot$ \citep{Fan2016a,Fan2018a,Tsai2015,SunWeibin2024}, heavily dust-obscured quasars \citep{Stern2014,Piconcelli2015,Vito2018}. Strong AGN rest frame UV / optical broad lines such as Ly$\alpha$, C\,{\sc iv}, Mg\,{\sc ii}, and H$\beta$ are found in their spectra \citep{Eisenhardt2012, Wu2012,Wu2018,Diaz2021}. \citet{Assef2016,Assef2020,Assef2022} studied a subsample of Hot DOGs that show stronger rest frame UV / optical emission than normal Hot DOGs and were called blue-excess Hot DOGs. Using the polarization imaging technique, \citet{Assef2022} have found that the rest-frame optical/UV excess emission is most likely scattered light from the central obscured quasar. \citet{Li2024} studied a larger sample of Hot DOGs, which contains both normal Hot DOGs and blue-excess Hot DOGs. They have found that the broad lines in normal Hot DOGs also originate from scattered light from the central obscured quasar, just as in blue-excess Hot DOGs.

Traditional methods for measuring the stellar mass of a galaxy include the $M/L$ method \citep{Bell2003} and the normal SED fitting method \citep{Bongiorno2007,Merloni2010,Suh2019}. In recent years, Bayesian SED fitting using Bayesian inference has been widely used as a more advanced method, which provides a more statistically robust and comprehensive estimate of parameters \citep[e.g.][]{Han2012,Han2014,Han2019,Han2023,Boquien2019,Yang2020}. Furthermore, in recent years, 2D image decomposition analysis based on high-resolution images has become a popular technique for better constraining stellar mass measurements \citep{Ding2020,LiJunyao2021a,LiJunyao2021b,Ding2023,LiJIH2021,LiJIH2023,Stone2024,Yue2024,Lijunyao2024,Sun2025,Tanaka2025,Yu2025}. To date, AGN studies using 2D image decomposition analyses have predominantly focused on type 1 AGNs or quasars. Moreover, to decompose the flux of the image into contributions from the extended host galaxy and the point-like AGN, most of the studies used moderate-luminosity samples. This is because if the luminosity of type 1 AGN/quasar is too high, the flux of the host will become too faint relative to the central AGN to be decomposed. Decomposing the host flux from hyperluminous $L_{bol} > 10^{13}L_\odot$ type 1 AGNs/quasars is currently extremely challenging. However, for Hot DOGs, the dust surrounding the central quasar allows only a small fraction of the UV/optical light of the quasar to reach us through scattering \citep{Assef2016,Assef2020,Assef2022}, making it possible for the host stellar component to be decomposed in the UV/optical band. Furthermore, the small fraction of scattered AGN broad-line components makes it possible for Hot DOGs to be detected by spectroscopic surveys. Previous studies have used the $M/L$ method and spectroscopic observations from large 8-10 m class telescopes to investigate the $M_{\rm BH}-M_\star$ relation of Hot DOGs \citep{Assef2015,Wu2018,Li2024}. In this work, we combine WISE photometric and SDSS spectroscopic survey data to select 11 hyperluminous Hot DOGs at $z=1.5-3.7$. We use a Bayesian SED decomposition method \citep{Han2012,Han2014,Han2019,Han2023} or combining image and SED decomposition together \citep{Yu2025} if high-resolution images such as \textit{Hubble Space Telescope} (HST) images are available to estimate the stellar mass. We combine the results of the decomposition with the emission line characteristics given by the SDSS to estimate the black hole mass. Then, the $M_{\rm BH}-M_\star$ relation of sources in the sample is studied. Most of the other works focus on the unobscured type I AGN, so the study of the $M_{\rm BH}-M_\star$ relation of hyperluminous dust-obscured quasars is the highlight of this work.

The paper is structured as follows. In Section \ref{sec:data}, we present the sample selection and multi-wavelength data description. In Section \ref{sec:method}, we describe the Bayesian SED decomposition method and the combined image and Bayesian SED decomposition method. The results and discussions are presented in Section \ref{sec:result} and Section \ref{sec:discussion}, respectively. We give a brief summary in Section \ref{sec:summary}. Throughout this work, we assume a flat ${\rm \Lambda}$CDM cosmology \citep{komatsu2011}, with $H_0=\rm{70\ km\ s^{-1}\ Mpc^{-1}}$, $\Omega_M = 0.3$ and $\Omega_\Lambda = 0.7$. Fluxes are corrected for Galactic extinction \citep{SF2011}.

\begin{deluxetable*}{cccccc}
\tablecaption{Properties of dust-obscured quasars selected with WISE and SDSS
\label{properties}}
\tablehead{
\colhead{Source $^{a}$} & \colhead{Redshift $^{b}$} & \colhead{log$M_\star$ $^{c}$} & \colhead{log$M_{\rm BH}$} & \colhead{log$L_{\rm bol}$} & \colhead{$\lambda_{\rm Edd}$} \\
\colhead{} & \colhead{} & \colhead{$(M_{\odot})$} & \colhead{$(M_{\odot})$} & \colhead{$(\mathrm{erg \ s^{-1}})$} & \colhead{}}
\startdata
SDSS J$150505.17+364916.8$ & 0.217 & $9.57^{+0.22}_{-0.25}$ & $7.36^{+0.60}_{-0.60}$ & 45.12 & 0.36 \\
SDSS J$112657.76+163912.0$ & 0.464 & $10.63^{+0.22}_{-0.18}$ & $8.63^{+0.31}_{-0.31}$ & 46.80 & 1.13 \\
SDSS J$163559.38+304032.8$ & 0.579 & $9.98^{+0.16}_{-0.15}$ & $8.34^{+0.38}_{-0.37}$ & 46.15 & 0.50 \\
\tableline
SDSS J$235511.59+070831.5$ & 1.543 & $11.16^{+0.15}_{-0.25}$ & $8.96^{+0.65}_{-0.65}$ & 47.24 & 1.05 \\
SDSS J$113931.08+460614.3$ & 1.820 & $...$ & $9.81^{+0.51}_{-0.51}$ & 46.92 & 0.10 \\
SDSS J$000521.65-085345.4$ & 2.300 & $11.12^{+0.24}_{-0.43}$ & $9.68^{+0.42}_{-0.42}$ & 47.36 & 0.37 \\
SDSS J$205122.47-004219.2$ & 2.450 & $11.08^{+0.14}_{-0.17}$ & $9.79^{+0.43}_{-0.43}$ & 47.83 & 0.79 \\
SDSS J$083448.48+015921.1$ & 2.594 & $9.97^{+0.24}_{-0.29}*$ & $9.28^{+0.41}_{-0.41}$ & 47.47 & 1.20 \\
SDSS J$085124.78+314855.7 \,\dagger$ & 2.638 & $10.67^{+0.13}_{-0.15}*$ & $9.08^{+0.41}_{-0.41}$ & 47.43 & 1.57 \\
SDSS J$135959.73+052512.3$ & 3.055 & $...$ & $9.80^{+0.54}_{-0.55}$ & 47.39 & 0.30 \\
SDSS J$022052.11+013711.1 \,\dagger$ & 3.138 & $11.17^{+0.07}_{-0.07}*$ & $9.08^{+0.42}_{-0.42}$ & 47.73 & 2.88 \\
SDSS J$011601.42-050503.9 \,\dagger$ & 3.183 & $11.28^{+0.18}_{-0.22}*$ & $9.14^{+0.43}_{-0.43}$ & 47.66 & 2.26 \\
SDSS J$015053.10-030528.7$ & 3.296 & $...$ & $9.04^{+0.41}_{-0.41}$ & 47.61 & 2.98 \\
SDSS J$101326.24+611219.7 \,\dagger$ & 3.703 & $11.73^{+0.21}_{-0.24}$ & $9.59^{+0.41}_{-0.41}$ & 47.76 & 0.82 \\
\enddata
\vspace{0.2cm}
\parbox{125mm} {
Notes. \\
The sources above and below the dividing line in the table belong to the relatively low-z and the relatively high-z subsamples, respectively. In this work, we focus on the relatively high-z subsample, in which the stellar masses of eight sources are estimated and used for subsequent analysis. \\
$^{a}$ Source names are from the \citet{Wu2022} SDSS quasar catalog. We use the abbreviation form in the text (e.g. SDSS J$022052.11+013711.1$ is abbreviated as J0220). Sources with three-component decomposition are marked with $\dagger$. $^{b}$ Redshifts are from the \citet{Wu2022} SDSS quasar catalog. $^{c}$ Stellar masses derived from the combining image and SED decomposition method are marked with *.}
\end{deluxetable*}

\begin{figure}
\epsscale{1.2}
\plotone{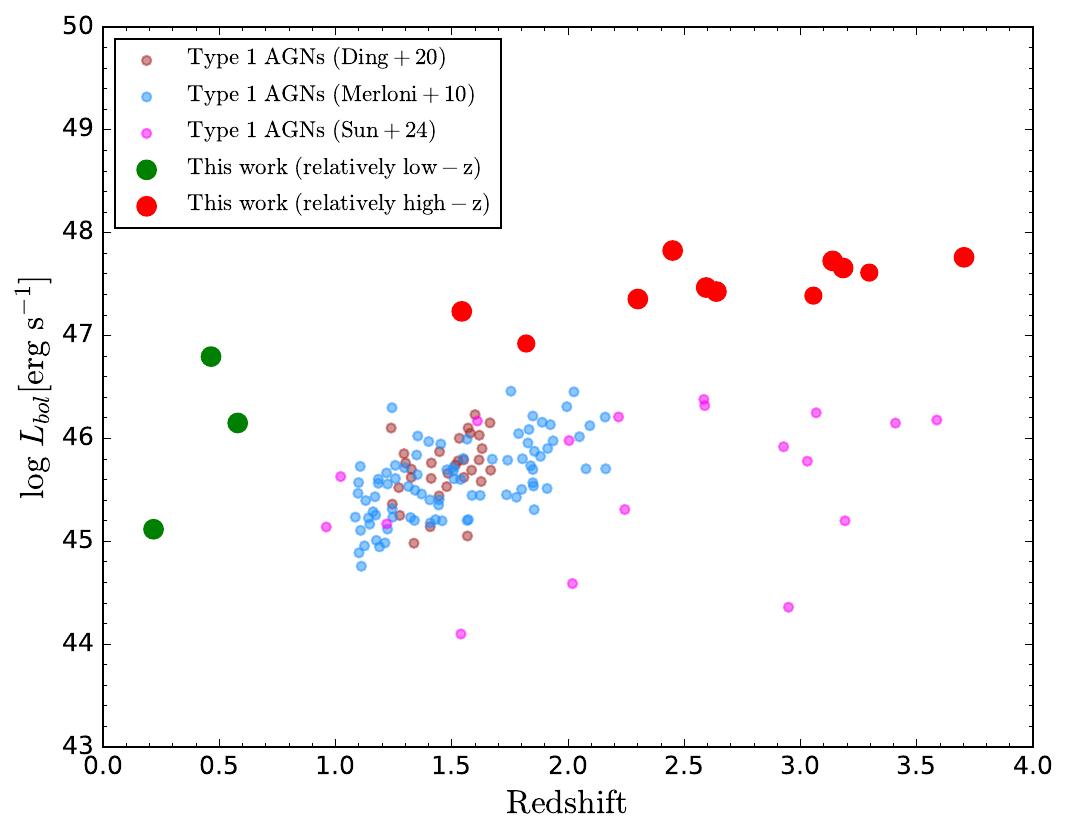}
\caption{The sources in our sample are distributed in the $L_{bol}-z$ plane. Sources in other works that study the $M_{\rm BH}-M_\star$ relation at $z=1-4$ are also shown in the figure. This figure shows that our Hot DOGs in the relatively high-z subsample are 1-2 dex brighter than sources at the same redshift range reported in other works, indicating that the $M_{\rm BH}-M_\star$ relation study of hyperluminous sources represents a key highlight of this work.}
\label{fig1}
\end{figure}

\section{Data}\label{sec:data}
\subsection{Sample Selection}\label{sec:sample}
We select Hot DOGs based on the WISE photometric and SDSS spectroscopic surveys data. We first select sources from ALLWISE Data Release \citep{Cutri2014} using the following selection criteria (magnitudes in Vega system):
\begin{align} \label{criteria}
\begin{split}
&\text{either} \\
&\quad \text{W4} < 7.7 \ \text{mag} \quad \text{and} \quad \text{W2} - \text{W4} > 8.2 \ \text{mag}, \\
&\text{or} \\
&\quad \text{W3} < 10.6 \ \text{mag} \quad \text{and} \quad \text{W2} - \text{W3} > 5.3 \ \text{mag}. \\
\end{split}
\end{align}
Compared to the ``W1W2-dropout" method used in \citet{Eisenhardt2012}, we remove the W1 $>$ 17.4 mag criterion. The W1 $>$ 17.4 mag criterion is used mainly to exclude background sources. However, since we have SDSS redshift data in this work, this criterion can be removed. Therefore, our selection criteria can be considered as the simplified Hot DOGs selection criteria relative to \citet{Eisenhardt2012}. The selected sources are then cross-matched with the catalog of quasar properties from the SDSS Data Release 16 \citep{Wu2022}, hereafter \citet{Wu2022} SDSS quasar catalog. 27 sources with broad AGN emission lines are selected after cross-matching. We perform a visual inspection of these 27 sources using WISE four-band images to eliminate spurious sources caused by blending or artifacts in the WISE images from our sample. Finally, we select 14 sources that pass the visual inspection, which are classified as Hot DOGs and used as the sample for the next analysis. We list source names and their redshift in Table \ref{properties}. We use the abbreviation form (e.g. SDSS J$022052.11+013711.1$ is abbreviated as J0220) of the names in the following text.

In Table \ref{properties}, we see that the galaxies can be divided into two subsamples based on redshift bins: a relatively low-z subsample with $z<0.6$ and a relatively high-z subsample with $z>1.4$. In this study, we focus on the relatively high-z subsample, specifically 11 Hot DOGs with z=1.5-3.7. We show the distribution of all sources in our sample on the bolometric luminosity versus redshift plane in Figure \ref{fig1}. The relatively low-z and high-z subsamples are represented by green and red dots, respectively. Figure \ref{fig1} also includes the results of previous studies. For the study of the $M_{\rm BH}-M_\star$ relation of high-redshift AGN, it is very important to cover the complete high redshift interval, so the previous work we selected here basically covers the high redshift interval of $1<z<4$. From Figure \ref{fig1}, we can see that in studies of the $M_{\rm BH}-M_\star$ relation, our Hot DOGs in the relatively high-z subsample are 1-2 dex brighter than the AGNs in other works in the same redshift range. The bolometric luminosities of these Hot DOGs are $L_{\rm bol} \gtrsim 10^{47}\,\mathrm{erg \ s^{-1}}$, which can be written as $L_{bol} > 10^{13}L_\odot$. These 11 Hot DOGs satisfy the definition of being hyperluminous. The study of these hyperluminous Hot DOGs will help us to understand the evolution of these extreme galaxies.

\subsection{Multi-wavelength Data}\label{sec:multiwavelength}
We assembled UV-to-millimeter data for our sample. All 14 sources have SDSS DR18 $ugriz$ bands photometry \citep{Almeida2023}. For $Y$, $J$, $H$, and $K$ bands, we cross matched the sample with surveys that has a detection in at least one band. We show the photometry of Hot DOGs in Table \ref{tab:Photometry_table}. The names of the surveys that contribute to $Y$, $J$, $H$, and $K$ bands can be found in the notes to Table \ref{tab:Photometry_table}. For WISE photometry from ALLWISE Data Release, 12 sources have all four bands photometry and two sources lack W4 band photometry. For FIR to millimeter data, three sources (J0851, J0220, and J0116) have \textit{Herschel} data \citep{Pilbratt2010} collected from the Herschel Science Archive (HSA)\footnote{\url{https://archives.esac.esa.int/hsa/whsa/index.html}}, the data including PACS \citep{Poglitsch2010} at 70 $\mu$m and 160 $\mu$m and SPIRE \citep{Griffin2010} at 250 $\mu$m, 350 $\mu$m and 500 $\mu$m. Three sources (J1126, J0220, and J0116) have ALMA data collected from ALMA Science Archive \footnote{\url{https://almascience.nrao.edu/aq/}}.
J1013 has HAWC+/SOFIA 89 $\mu$m and 154 $\mu$m, SCUBA-2/JCMT 450 $\mu$m and 850 $\mu$m, SMA 870 $\mu$m and 1.3mm data from \citet{Toba2020}.

Four sources (J0851, J0220, J0116, and J0834) have HST imaging from MAST \footnote{\url{https://mast.stsci.edu/portal/Mashup/Clients/Mast/Portal.html}}. These high-resolution images are used in the decomposition in Section \ref{sec:combining}.

\section{Methods} \label{sec:method}
\subsection{SED Analysis}\label{sec:SED}
The SED analysis from UV to millimeter is performed using the Bayesian SED fitting code BayeSED3 \citep{Han2012,Han2014,Han2019,Han2023}. We model the stellar emission by adopting the \citet{Bruzual2003} simple stellar population (SSP). We assume the \citet{Chabrier2003} initial mass function (IMF), the \citet{Calzetti2000} dust attenuation law, and an exponentially declining star formation history (SFH). We model the AGN emission by adopting the CLUMPY model \citep{Nenkova2008a,Nenkova2008b}. The CLUMPY model includes the torus dust emission, and a part of AGN emission scattered into our line of sight or not absorbed by torus dust. The model has been used to fit SEDs of type II Seyferts with scattered light that includes broad line components \citep{Ichikawa2015}. This suggests that the CLUMPY model is suitable for modeling Hot DOGs, especially blue-excess Hot DOGs. For sources that have FIR to millimeter data, the SEDs are decomposed into three components: stellar, AGN, and cold dust. The cold dust emission results from a re-emitted process in which the energy of stellar emission absorbed by dust is assumed to be totally re-emitted at the IR band. The cold dust emission is modeled as a graybody $S_{\lambda}\propto(1-e^{-(\frac{\lambda_0}{\lambda})^{\beta}}) B_\lambda(T_{dust})$, where $\lambda_0$ = 125$\mu$m, $B_\lambda$ is the Planck blackbody spectrum. The emissivity index $\beta$ and dust temperature $T_{\rm dust}$ are two free parameters. The stellar masses ($M_\star$) and the bolometric luminosities ($L_{\rm bol}$) derived from SED analysis are shown in Table \ref{properties}. For the 11 Hot DOGs in the relatively high-z subsample, we derive the stellar masses for eight sources. The details can be found in Section \ref{sec:result}.

\begin{figure*}
\plotone{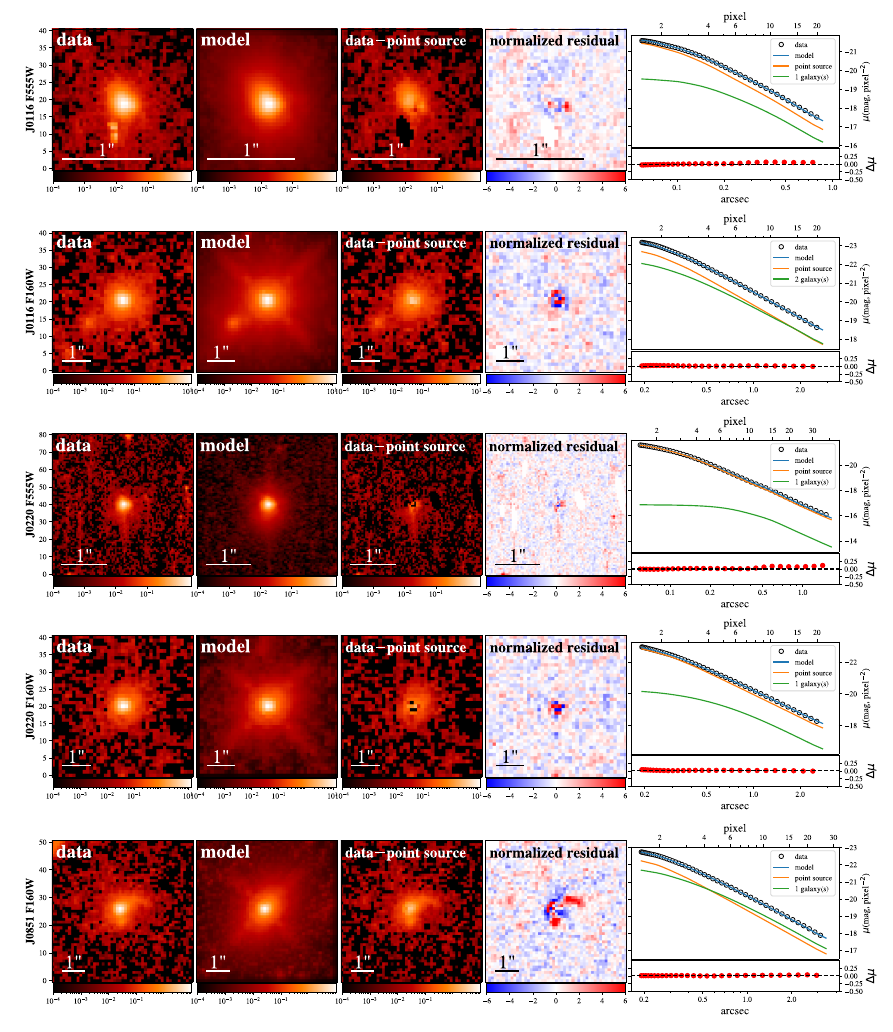}
\caption{
Best-fit image decomposition results for the three sources (J0116, J0220 and J0851) obtained using \texttt{GaLight}. The panels from left to right are: (1) observed data, (2) best-fit Sersic + point source model, (3) observed data minus the point source model (4) residual divided by variance and (5) radial surface brightness
profile (top) and residual (bottom).
This profile includes the data (open circles), best-fit model (blue curve), the point source model of AGN (orange curve) and the model of the host galaxy (green curve). The fitting is based on 2-dimensional image, while the
1-dimensional profile is only an illustration of the fitting result.
}
\label{fig2}
\end{figure*}

\subsection{Combining image and SED decomposition}\label{sec:combining}
Among the 14 sources in our sample, we notice that 4 have corresponding HST images from previous work \citep{Fan2016b,Zakamska2019,Assef2020}. Following the method from \citet{Yu2025}, we incorporate morphology constraint from image decomposition into SED fitting. Using \texttt{GaLight} \citep{Ding2021}, we simultaneously fit the host galaxy and the AGN as a Sersic profile and a point source. For each source, we find all isolated and non-saturated sources in the field of view and reconstruct a median stack PSF using PSFr (Birrer et al. in prep). PSFr is a python software for constructing PSF. PSFr calculates sub-pixel astrometric shifts, making the resulting PSF highly accurate. The uncertainty in the flux of the fitted components is estimated to be 20\% percent of the fitted flux, which could originate from various sources such as inaccurate PSF profiles used in the decomposition \citep[e.g.][]{Tanaka2025}. For image decomposition, in practice, when the flux of the host galaxy is more than an order of magnitude lower than that of the central AGN, the radiation from the central AGN completely overwhelms that of the host galaxy, making the results of image decomposition relatively unreliable. This is the primary reason why image decomposition studies are rarely conducted for Type 1 quasars. We show image decomposition results of three sources (J0116, J0220, and J0851) in Figure \ref{fig2}. For these three sources, the difference in flux between the host galaxy and the AGN is within an order of magnitude. Specifically, the flux of the host galaxies of J0116 and J0851 is comparable to that of the AGN. This makes the image decomposition results relatively reliable. In contrast, for J0834, the flux of the decomposed host galaxy in the F814W and F160W images is more than an order of magnitude lower than that of the central AGN, making the decomposition of the host galaxy exceedingly challenging due to the overwhelming contribution of the AGN. Consequently, the reliability of the image decomposition results for J0834 is significantly lower compared to those for J0116, J0220, and J0851. Therefore, J0834 is not shown in Figure \ref{fig2}.

\begin{figure}
\plotone{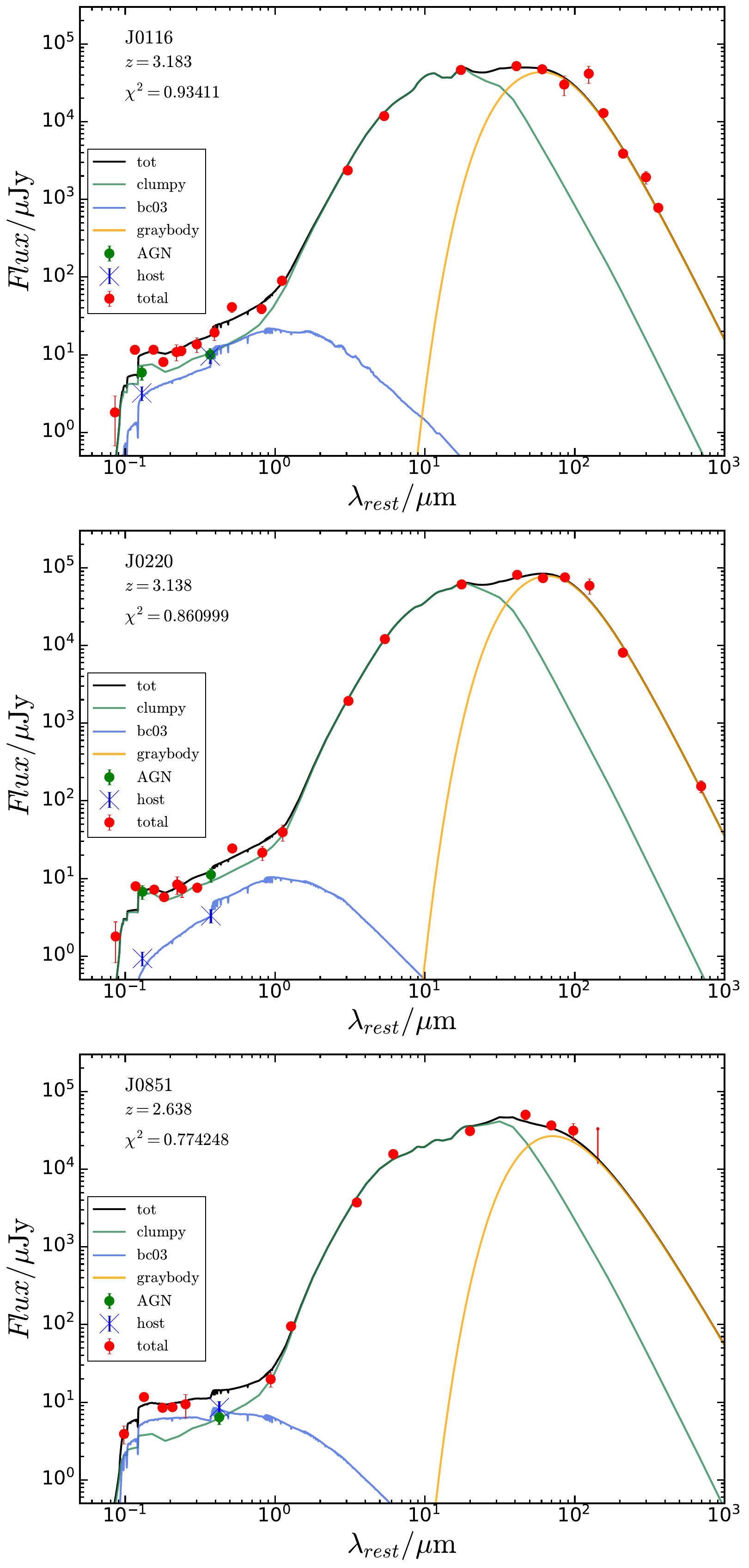}
\caption{Three components best-fit SEDs of three sources which use the combining image and SED decomposition method. The red points represent the observed data. The blue crosses and green points are the fluxes of stellar and AGN component derived from image decomposition. The green, blue, and orange solid lines represent the emissions from AGN, stellar, and cold dust components, respectively. The black solid line represents the total model fit.
\label{fig3}}
\end{figure}

In the SED fitting procedure in Section \ref{sec:SED}, we define a new form of likelihood function to additionally constrain the galaxy (AGN) model with the flux obtained through image decomposition, as presented in \citet{Yu2025}. The results of the combination of the image and the SED decomposition of three sources are shown in Figure \ref{fig3}.

\subsection{Black hole mass estimation}\label{sec:MBH}
\citet{Wu2022} SDSS quasar catalog is based on the SDSS Data Release 16 quasar catalog \citep{Lyke2020} and provides further analysis of the properties of quasars. In the \citet{Wu2022} SDSS quasar catalog, BH masses are estimated from measurements of the continuum and broad emission lines \citep{Wu2022}. Estimates are based on the ``single-epoch virial BH mass" estimators \citep{Vestergaard2006}. We show the fiducial BH mass recipes on H$\beta$, Mg\,{\sc ii} and C\,{\sc iv} adopted in \citet{Wu2022} here.
\begin{equation}\label{MBH_estimator}
\log\left(\frac{M_{\mathrm{BH}}}{M_{\odot}}\right) = a + b \log\left(\frac{\lambda L_{\lambda}}{10^{44} \, \mathrm{erg \, s^{-1}}}\right) + 2 \log\left(\frac{\mathrm{FWHM}}{\mathrm{km \, s^{-1}}}\right),
\end{equation}
BH mass estimates of sources at different redshifts rely on different broad emission lines. where $(a, b)=(0.910, 0.50)$ for H$\beta$, $(0.740, 0.62)$ for Mg\,{\sc ii}, and $(0.660, 0.53)$ for C\,{\sc iv}. FWHM is the full width at half-maximum of the broad emission line. $\lambda L_{\lambda}$ is the monochromatic continuum luminosity at 5100, 3000, and 1350 \AA, corresponds to H$\beta$, Mg\,{\sc ii}, and C\,{\sc iv}.

However, due to the high dust obscuration of our sources in the UV/optical band, we cannot use the $\lambda L_{\lambda}$ values of the \citet{Wu2022} SDSS quasar catalog, which are measured directly from SDSS spectra and are used mainly for Type 1 quasars. Instead, we use the CLUMPY model \citep{Nenkova2008a,Nenkova2008b} adopted in the SED fitting to derive the obscuration-corrected monochromatic continuum luminosity. Then we substitute the FWHM of the broad emission line from the \citet{Wu2022} SDSS quasar catalog and the obscuration-corrected monochromatic continuum luminosity into equation \ref{MBH_estimator} to derive BH masses of our sources, the BH masses value are shown in Table \ref{properties}.

\section{Results} \label{sec:result}
For the 11 Hot DOGs in the relatively high-z subsample, the stellar masses and their confidence intervals of eight sources are estimated by combining images and Bayesian SED decomposition or by Bayesian SED decomposition alone. For the remaining three sources in the relatively high-z subsample, due to insufficient data in the UV-optical band and lack of FIR-millimeter data, SED decomposition cannot estimate the stellar mass and its confidence interval. The ellipses (...) in Table \ref{properties} indicate the sources that lacked a stellar mass estimate. In Figure \ref{fig3}, we show the best-fit SEDs of three sources with both FIR-millimeter observations and high-resolution images from the HST using the combining image and SED decomposition method. For these three sources (J0116, J0220, and J0851), which have the most complete data, the stellar mass of galaxies can be estimated more accurately using the combined image and SED decomposition.

It is worth noting that J0220 and J0116 have been identified as blue-excess Hot DOGs in previous stuides, where the rest frame UV/optical flux is dominated by AGN emission \citep{Assef2016,Assef2020}. \citet{Assef2022} used VLT/FORS2 to perform imaging polarimetry of J0116 in the $R$ band, and found that the rest-frame UV/optical flux of J0116 is strongly linearly polarized, indicating that the flux is most likely scattered light from the central obscured quasar. In Figure \ref{fig3}, we see that the rest-frame UV/optical flux of these two sources is dominated by AGN emission (as indicated in Section \ref{sec:SED}, the CLUMPY model represents scattered light in rest-frame UV/optical). This finding is consistent with the imaging polarimetry observations of J0116 \citep{Assef2022}, suggesting that the CLUMPY model can describe the physical properties of these sources well. J0851 has been identified as a Hot DOG in \citet{Wu2012}. From the bottom panel of Figure \ref{fig3}, we find that its AGN emission does not dominate the rest-frame UV/optical flux. Therefore, we identify this source as a normal Hot DOG.

For three sources at relatively low redshifts, J1126 at $z=0.464$ has $L_{bol} > 10^{13}L_\odot$ and may satisfy the definition of low-redshift Hot DOGs in \citet{Li2023,Li2025}. The low-redshift Hot DOGs are very similar to the high-redshift Hot DOGs in the SED shape. We will focus on J1126 in future work (Luo et al. in preparation).

\section{Discussion} \label{sec:discussion}
For sources in our sample, we calculate the Eddington ratio $\lambda_{\rm Edd}=L_{\rm AGN}/L_{\rm Edd}$, $L_{\rm AGN}$ is the bolometric luminosity of the AGN obtained by integrating the AGN component. The derived properties are listed in Table \ref{properties}. For the 11 Hot DOGs in the relatively high-z subsample, their Eddington ratios are generally high, with a median of 1.05, which is consistent with previous studies of Hot DOGs \citep{Wu2018,Tsai2018,Li2024}. We found that the Eddington ratios of these Hot DOGs are similar to those of $z\sim 6$ quasars \citep{Wang2010,Yue2024,Loiacono2024}. Research has found that $z\sim 6$ quasars are often located in high-density environments, experiencing frequent mergers and having abundant gas supply, which is responsible for their high Eddington ratios \citep{Morselli2014,Decarli2024,Loiacono2024,Trinca2024}. Given that Hot DOGs are also frequently found in high-density regions with high merger rates \citep{Fan2016b,Jones2014,Diaz2018,Luo2022,Luo2024,Zewdie2023}, we propose that this may explain their high Eddington ratios.

In addition, the FIR-millimeter data can well constrain the SFR in the SED fitting. For the four sources (J0851, J0220, J0116, and J1013) with available FIR-millimeter data, the SFRs are 453.85, 1616.40, 938.66, and 2626.76 $M_{\odot} \ \rm yr^{-1}$, respectively, which suggests that there is intense star-formation activity in these galaxies. Therefore, there are both intense star formation and rapid black hole accretion in these Hot DOGs, supporting the perspective that Hot DOGs are in the hybrid phase of starbursts and AGN activity \citep{Fan2016a,SunWeibin2024}.

\begin{figure*}
\plotone{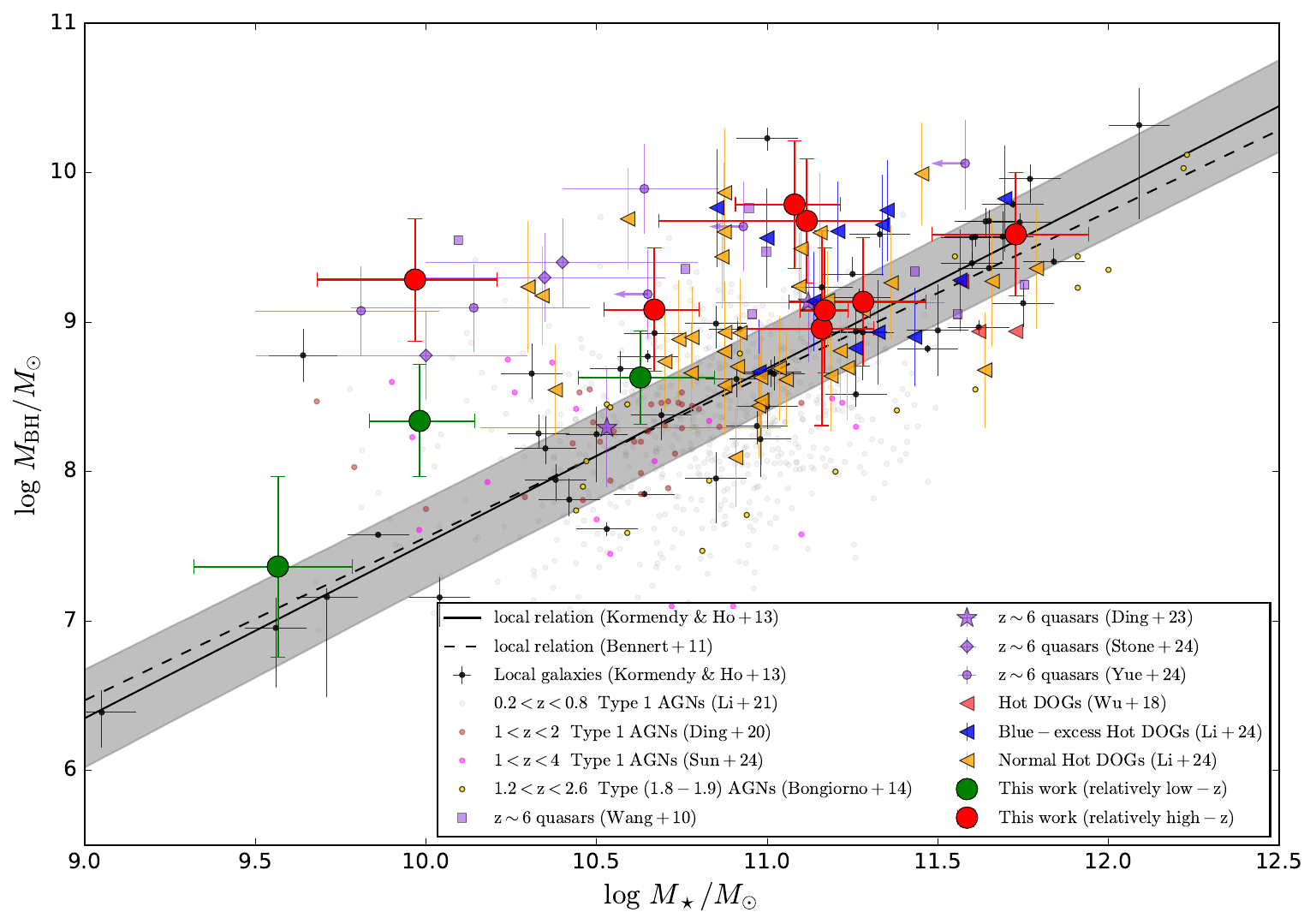}
\caption{
$M_{\rm BH}-M_\star$ relation. The green and red circles represent relatively low redshift and high redshift sources in our sample, respectively. The local relation and local galaxies from \citet{Kormendy2013} are shown using solid line and black dots. The gray filled area represent the $1 \sigma$ scatter of the local relation from \citet{Kormendy2013}. The dashed line represent the local relation from \citet{Bennert2011}. Type 1 AGNs within the redshift range of $0.2 < z < 4$  \citep{LiJunyao2021b,Ding2020,Sun2025}, Type (1.8-1.9) AGNs within the redshift range of $1.2 < z < 2.6$ from \citep{Bongiorno2014}, and $z\sim 6$ quasars \citep{Wang2010,Ding2023,Stone2024,Yue2024} are shown. Previous $M_{\rm BH}-M_\star$ relation studies of Hot DOGs using $M/L$ method are also plotted, Hot DOGs in \citet{Wu2018} are shown as red left triangles. Blue-excess and normal Hot DOGs in \citet{Li2024} are shown as green and orange left triangles, respectively. This figure indicates that there are no significant differences in the $M_{\rm BH}-M_\star$ relation of Hot DOGs compared to the local relation.}
\label{fig4}
\end{figure*}

In Figure \ref{fig4}, we plot the black hole masses and stellar masses of sources in our sample on the log scale $M_{\rm BH}-M_\star$ diagram to investigate the $M_{\rm BH}-M_\star$ relation. We include the local relation and its scatter range from \citet{Kormendy2013} in the diagram, along with the local galaxies used in their work. Type 1 AGNs within the redshift range of $0.2 < z < 4$ from \citep{LiJunyao2021b,Ding2020,Sun2025} and type 1.8-1.9 AGNs within the redshift range of $1.2 < z < 2.6$ from \citep{Bongiorno2014} are also plotted. For AGNs in the early universe, we plot the results of the work of $z\sim 6$ quasars \citep{Wang2010,Ding2023,Stone2024,Yue2024}. In addition, we also show the values of $M_{\rm BH}$ and $M_\star$ of Hot DOGs estimated from previous studies \citep{Wu2018,Li2024}. \citet{Wu2018} used the maximum value of the $M/L$ constraint range to estimate the upper limits of the stellar masses. \citet{Li2024} used the relation between the optical color of the rest frame and the $M/L$ ratio to estimate the stellar masses. Noting that in these Hot DOGs studies, a local relation from \citet{Bennert2011} is used, which is very close to the local relation from \citet{Kormendy2013}. The results of Hot DOGs compared to \citet{Bennert2011} local relation can be seen in Figure 9 in \citet{Wu2018}.

\begin{figure}
\plotone{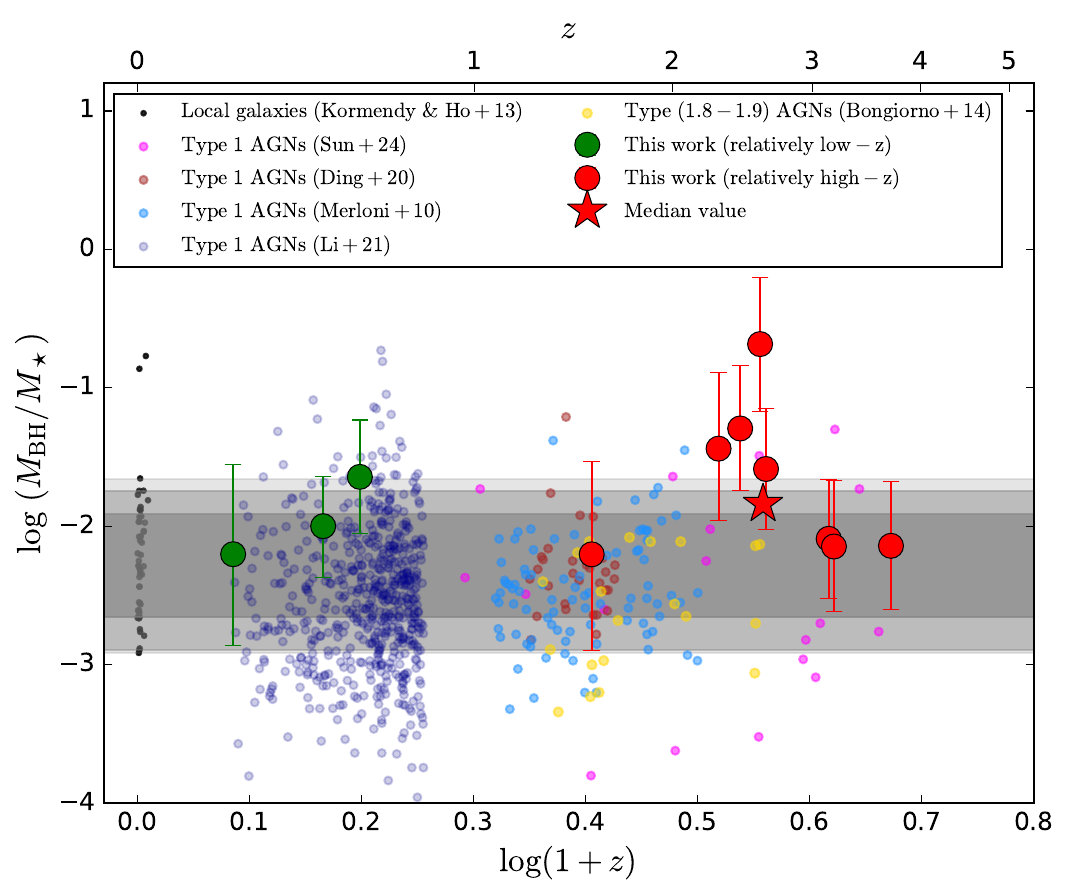}
\caption{The red star marked the median log $(M_{\rm BH}/M_\star)$ of the eight Hot DOGs. log $(M_{\rm BH}/M_\star)$ from redshift $0-4$ in different population are also shown. The gray filled areas of different opacity represent the scatter ranges of log $(M_{\rm BH}/M_{\rm bulge})$ for local galaxies at 1, 2, and 3 $\sigma$ levels, respectively \citep{Kormendy2013}.
\label{fig5}}
\end{figure}

We find that among the eight Hot DOGs, four reside within the $M_{\rm BH}-M_{\rm bulge}$ relation in the local universe, while the remaining four sources lie above this local relation. There is significant scatter in our sample. To better compare Hot DOGs with the local relation, we show log $M_{\rm BH}/M_\star$ as a function of the redshift in Figure \ref{fig5}. For log $(M_{\rm BH}/M_\star)$ of the eight Hot DOGs, the 16th, 50th, and 84th percentiles are $-2.14$, $-1.83$, and $-1.31$, respectively. The red star marked the median log $(M_{\rm BH}/M_\star)$ of the eight Hot DOGs. Consider that two out of the three sources in our sample with the most complete data are located within the local relation, and one is above it. In addition, J1013, which also has relatively complete multi-wavelength data, is situated within the local relation. The insufficient multi-wavelength data for some Hot DOGs result in larger estimation errors for $M_{\rm BH}$ and $M_\star$, which may contribute to the significant scatter observed in these eight Hot DOGs. However, this large scatter might also indicate that there may exist intrinsic scatter within the population of Hot DOGs. It is worth noting that the Hot DOG from \citet{Li2024} also show large scatter, with some sources within the local relation and other above the local relation. Considering previous studies that have found the Hot DOGs stage is triggered and sustained by multiple minor mergers (small satellite galaxies merging into the primary galaxy) \citep{Diaz2021,Ginolfi2022,Li2024}, we hypothesize that minor mergers may be the intrinsic mechanism driving the observed scatter. Minor mergers can significantly increase stellar mass without significantly increasing black hole mass.

Four Hot DOGs (J2355, J0220, J0116, and J1013) lie within the $1 \sigma$ scatter range of log $(M_{\rm BH}/M_{\rm bulge})$ observed in the local universe, whereas the remaining sources (J0005, J2051, J0834, and J0851) lie above this range. Among these Hot DOGs above this range, J0851 has the lowest deviation, while J0834 has the largest deviation. For J0834, this is primarily due to fitting issues, as the photometric data of J0834 in the rest-frame wavelength range of $\sim 2400-2700 $~\AA\ exhibits an anomalous bump, as can be seen from Table \ref{tab:Photometry_table}. This phenomenon has been observed in survey data from SDSS, Pan-STARRS1, and HSC. We check the images and preliminarily determine that this is not caused by photometric contamination. It may be an intrinsic feature of the galaxy, but existing AGN and stellar templates fail to adequately explain this phenomenon, leading to poor fitting results. Consequently, the resulting ratio, although retained in the figure, is noted to have low reliability. We prepare to conduct a detailed investigation of this phenomenon in future work (Luo et al. in preparation).

Even if we include J0834, the median log $(M_{\rm BH}/M_\star)$ of the eight sources is only slightly higher than the $1 \sigma$ scatter range of log $(M_{\rm BH}/M_{\rm bulge})$ in the local universe, but it still falls within the $2 \sigma$ scatter range. This suggests that Hot DOGs, as a whole population, its' $M_{\rm BH}-M_\star$ relation may not have significant differences relative to the local relation.

In the framework of galaxy formation and evolution, starburst galaxies, dust-obscured quasars, optically bright quasars, and massive early-type galaxies are considered to be in an evolutionary sequence, and dust-obscured quasars are considered to be the progenitor of massive elliptical galaxies in the local universe \citep{Hopkins2008,Alexander2012}. For starburst galaxies such as sub-millimeter galaxies (SMGs), the estimation of black hole masses is primarily based on assuming a fixed Eddington ratios due to the severe obscuration that makes it difficult to observe broad emission lines \citep{Borys2005,Alexander2008}, and these studies indicate that the SMGs are located below the local relation. \citet{Zhuang2023} suggests that starburst galaxies below the local relation will evolve more in the vertical direction with significant $M_{\rm BH}$ growth and finally return to the local relation of early-type galaxies. Hot DOGs represent a critical stage in this evolutionary sequence. The presence of heavy obscuration and scattered AGN light in these sources allows us to estimate both stellar masses and black hole masses simultaneously. We find that these sources deviate only slightly from the local relation, and their black hole masses have reached $10^{9-10}M_{\odot}$, comparable to massive early-type galaxies in the local universe. This indicates that if these sources are progenitors of massive early-type galaxies, their mass assembly process has been largely completed. If these sources remain on the local relation, this suggests that their evolution in the $M_{\rm BH}/M_\star$ diagram is minimal. Considering previous studies have found AGN feedback such as ionized gas or molecular gas outflows in some Hot DOGs \citep{Diaz2016,Fan2018b,Finnerty2020,Jun2020}. We speculate that for Hot DOGs, although star formation and black hole accretion are still strong at the current stage, the strong AGN feedback will ``quench" them in a very short period of time, so that these sources will remain on the local relation.

To gain more insight into the potential evolutionary sequence of Hot DOGs. We compare Hot DOGs with another hyperluminous population, the WISSH quasars \citep{Duras2017}, whose bolometric luminosity $L_{\rm bol} \gtrsim 10^{47}\,\mathrm{erg \ s^{-1}}$ . \citet{Duras2017} have found that the WISSH quasars are unobscured quasars, but their SFRs are high, up to $\sim 2000 \ M_{\odot} \ \rm yr^{-1}$. Similar bolometric luminosity and SFR suggest that there may be an evolutionary link between the two populations, although further research is required for a detailed connection.

\section{Summary and conclusion}\label{sec:summary}
In this work, we select 11 hyperluminous Hot DOGs at $z=1.5-3.7$ based on the ALLWISE Data Release and the \citet{Wu2022} SDSS quasar catalog. BH masses of these sources are estimated via ``single-epoch virial BH mass" estimators. We derive the stellar masses of these sources via either Bayesian SED fitting or a combination of HST image and SED decomposition. For these 11 hyperluminous Hot DOGs,stellar masses were successfully measured from eight sources. We find high Eddington ratios in our sample, with the median value of 1.05 and the maximum value close to 3. Hot DOGs are often found in overdense environments. Frequent mergers and having abundant gas supply in overdense environments may be the reason for super-Eddington accretion.

We find that the log $(M_{\rm BH}/M_\star)$ of these Hot DOGs shows a large scatter, which originates primarily from measurement errors and intrinsic scatter within this population. However, the median log $(M_{\rm BH}/M_\star)$ of these Hot DOGs is only slightly higher than the $1 \sigma$ scatter range of log $(M_{\rm BH}/M_{\rm bulge})$ in the local universe and half of the Hot DOGs lie on the local relation. This suggests that as a whole population, no significant evolution of the $M_{\rm BH}-M_\star$ relation of these Hot DOGs compared to the local relation. This indicates that Hot DOGs may be the progenitor of massive early-type galaxies in the local universe. We speculate that the subsequent evolution of these sources may be significantly influenced by AGN feedback and remain on the local relation.

\begin{acknowledgments}
We express our sincere gratitude to the anonymous referee for their valuable and insightful comments. The authors sincerely thank Dr. Guodong Li for the valuable discussions via email. This work is supported by National Key Research and Development Program of China (2023YFA1608100). The authors gratefully acknowledge the support provided by the National Natural Science Foundation of China (NSFC, Grant Nos. 12173037, 12233008), the CAS Project for Young Scientists in Basic Research (No. YSBR-092), the Fundamental Research Funds for the Central Universities (WK3440000006), Cyrus Chun Ying Tang Foundations and the 111 Project for ``Observational and Theoretical Research on Dark Matter and Dark Energy'' (B23042).  YH acknowledges support from the National Science Foundation of China (grant Nos. 11773063), the Light of West China program of the CAS, and the Yunnan Ten Thousand Talents Plan Young \& Elite Talents Project.
\end{acknowledgments}

\appendix
\section{Appendix information}
Here we include a supplementary table that shows the photometry of Hot DOGs.

\tabletypesize{\footnotesize}
\begin{deluxetable}{cccccccccccccc}
\rotate
\tablecaption{Photometry of Hot DOGs \label{tab:Photometry_table}}
\tablehead{
Source & $u$ & $g$ & $r$ & $i$ & $z$ & $Y$ & $J$ & $H$ & $K$ & W1 & W2 & W3 & W4 \\
\colhead{} & \colhead{($\mu$Jy)} & \colhead{($\mu$Jy)} & \colhead{($\mu$Jy)} & \colhead{($\mu$Jy)}  & \colhead{($\mu$Jy)}  & \colhead{($\mu$Jy)}  & \colhead{($\mu$Jy)}  & \colhead{($\mu$Jy)}  & \colhead{($\mu$Jy)}  & \colhead{($\mu$Jy)}  & \colhead{($\mu$Jy)}  & \colhead{($\mu$Jy)}  & \colhead{($\mu$Jy)} 
}
\startdata
J1505 & $32.0 \pm 1.5$ & $39.1 \pm 0.6$ & $53.5 \pm 0.8$ & $70.4 \pm 1.2$ & $75.7 \pm 4.0$ & $69.7 \pm 3.3$ $^{a}$ & $99.8 \pm 4.0$ $^{c}$ & $...$ & $167.7 \pm 8.2$ $^{c}$ & $121.6 \pm 4.9$ & $153.2 \pm 8.0$ & $2276.2 \pm 115.3$ & $15645.9 \pm 749.3$ \\
J1126 & $58.2 \pm 2.6$ & $82.5 \pm 1.1$ & $170.2 \pm 1.7$ & $232.0 \pm 2.4$ & $286.6 \pm 9.0$ & $194.9 \pm 3.4$ $^{a}$ & $215.8 \pm 6.4$ $^{c}$ & $...$ & $453.5 \pm 7.5$ $^{c}$ & $648.0 \pm 17.3$ & $785.2 \pm 21.7$ & $17636.0 \pm 341.1$ & $98900.8 \pm 2550.6$ \\
J1635 & $11.2 \pm 1.3$ & $14.3 \pm 0.6$ & $26.1 \pm 0.8$ & $39.9 \pm 1.3$ & $43.6 \pm 4.1$ & $46.4 \pm 3.0$ $^{a}$ & $39.6 \pm 2.7$ $^{d}$ & $57.7 \pm 5.7$ $^{d}$ & $80.7 \pm 5.9$ $^{d}$ & $209.3 \pm 26.2$ & $280.2 \pm 10.1$ & $4843.9 \pm 178.5$ & $27189.4 \pm 1252.1$ \\
J2355 & $13.3 \pm 1.7$ & $18.2 \pm 0.6$ & $27.4 \pm 0.8$ & $38.8 \pm 1.1$ & $48.4 \pm 3.8$ & $39.5 \pm 4.3$ $^{a}$ & $50.5 \pm 6.2$ $^{d}$ & $70.0 \pm 5.5$ $^{d}$ & $63.3 \pm 6.5$ $^{d}$ & $119.3 \pm 6.8$ & $141.8 \pm 11.9$ & $4073.8 \pm 180.1$ & $...$ \\
J1139 & $8.0 \pm 1.3$ & $6.1 \pm 0.5$ & $8.0 \pm 0.9$ & $7.9 \pm 1.4$ & $7.1 \pm 4.5$ & $...$ & $...$ & $...$ & $...$ & $45.5 \pm 4.1$ & $92.1 \pm 8.2$ & $2932.2 \pm 124.2$ & $9315.4 \pm 866.6$ \\
J0005 & $4.7 \pm 1.5$ & $7.5 \pm 0.5$ & $9.3 \pm 0.8$ & $9.0 \pm 1.3$ & $15.7 \pm 4.8$ & $...$ & $...$ & $...$ & $...$ & $50.6 \pm 5.3$ & $69.9 \pm 13.2$ & $1261.2 \pm 169.6$ & $8995.0 \pm 1035.6$ \\
J2051 & $28.9 \pm 1.9$ & $45.7 \pm 0.8$ & $43.1 \pm 0.8$ & $45.1 \pm 1.1$ & $61.9 \pm 3.8$ & $71.7 \pm 2.5$ $^{a}$ & $52.1 \pm 4.9$ $^{d}$ & $61.0 \pm 6.8$ $^{d}$ & $68.0 \pm 6.1$ $^{d}$ & $52.1 \pm 5.8$ & $60.8 \pm 13.2$ & $3694.9 \pm 343.7$ & $...$ \\
J0834 & $3.5 \pm 0.9$ & $22.8 \pm 0.6$ & $13.0 \pm 0.6$ & $15.5 \pm 1.0$ & $45.5 \pm 3.7$ & $49.9 \pm 2.9$ $^{a}$ & $21.1 \pm 1.9$ $^{e}$ & $36.8 \pm 2.8$ $^{e}$ & $24.0 \pm 3.2$ $^{e}$ & $44.6 \pm 6.6$ & $106.2 \pm 11.4$ & $4373.2 \pm 217.5$ & $14401.2 \pm 1246.8$ \\
J0851 & $3.9 \pm 1.0$ & $11.7 \pm 0.5$ & $8.5 \pm 0.6$ & $8.6 \pm 0.8$ & $9.4 \pm 3.1$ & $...$ & $...$ & $...$ & $...$ & $19.8 \pm 4.2$ & $95.0 \pm 12.3$ & $3729.1 \pm 182.0$ & $15660.3 \pm 1096.2$ \\
J1359 & $0.5 \pm 0.6$ & $9.4 \pm 0.6$ & $5.5 \pm 0.6$ & $7.7 \pm 1.0$ & $11.0 \pm 4.1$ & $...$ & $...$ & $...$ & $...$ & $27.9 \pm 4.1$ & $109.3 \pm 9.6$ & $3317.4 \pm 143.6$ & $10083.2 \pm 845.1$ \\
J0220 & $1.8 \pm 1.0$ & $7.9 \pm 0.4$ & $7.2 \pm 0.6$ & $5.8 \pm 0.7$ & $8.4 \pm 2.1$ & $7.4 \pm 1.7$ $^{b}$ & $7.6 \pm 1.0$ $^{g}$ & $...$ & $24.4 \pm 2.6$ $^{g}$ & $21.5 \pm 4.5$ & $39.5 \pm 9.2$ & $1930.2 \pm 133.3$ & $12067.0 \pm 1022.5$ \\
J0116 & $1.8 \pm 1.1$ & $11.6 \pm 0.6$ & $11.6 \pm 0.6$ & $8.1 \pm 0.7$ & $10.8 \pm 2.6$ & $11.2 \pm 1.6$ $^{b}$ & $13.6 \pm 2.9$ $^{f}$ & $19.4 \pm 4.1$ $^{f}$ & $41.1 \pm 5.9$ $^{f}$ & $38.9 \pm 5.4$ & $89.9 \pm 12.0$ & $2361.6 \pm 152.3$ & $11814.1 \pm 1109.9$ \\
J0150 & $0.2 \pm 0.1$ & $4.8 \pm 0.4$ & $7.2 \pm 0.6$ & $5.8 \pm 0.8$ & $5.0 \pm 2.4$ & $2.5 \pm 1.3$ $^{b}$ & $8.6 \pm 2.9$ $^{f}$ & $...$ & $28.1 \pm 7.0$ $^{f}$ & $30.4 \pm 4.8$ & $32.4 \pm 10.6$ & $1617.3 \pm 151.9$ & $8847.1 \pm 1181.5$ \\
J1013 & $1.6 \pm 0.8$ & $3.6 \pm 0.5$ & $14.0 \pm 0.7$ & $13.8 \pm 1.0$ & $21.3 \pm 3.9$ & $...$ & $...$ & $...$ & $...$ & $45.6 \pm 7.0$ & $134.3 \pm 9.0$ & $3296.1 \pm 163.9$ & $10695.5 \pm 985.1$ \\
\enddata
\vspace{0.2cm}
\parbox{\textwidth} {
Notes. \\
All 14 sources have SDSS DR18 $ugriz$ bands photometry \citep{Almeida2023}. \\
$^{a}$ Pan-STARRS1 survey (PS1) DR2 \citep{Chambers2016,Flewelling2020}. \\
$^{b}$ Dark Energy Survey (DES) DR2 \citep{Abbott2021}. \\
$^{c}$ UK Infra-Red Telescope (UKIRT) Hemisphere Survey (UHS) DR2 \citep{Dye2018}. \\
$^{d}$ UKIRT Infrared Deep Sky Survey (UKIDSS) Large Area Survey (LAS) DR11 \citep{Lawrence2007}. \\
$^{e}$ VISTA Kilo-Degree Infrared Galaxy Survey (VIKING) DR5 \citep{Edge2013}. \\
$^{f}$ VISTA Hemisphere Survey (VHS) DR6 \citep{McMahon2013}. \\
$^{g}$ Photometry from \citet{Assef2020}. \\
}

\end{deluxetable}

\bibliography{ms}{}
\bibliographystyle{aasjournal}

\end{CJK*}
\end{document}